\documentclass[twocolumn]{aa}
\usepackage{graphicx}
\usepackage{txfonts}
\begin{document}
   \title{Spiral shock detection on eclipse maps: \\
 Simulations and Observations\thanks{Based on
observations made  with    the  Isaac Newton  and   William   Herschel
telescopes operated  on the island  of  La Palma  by the Isaac  Newton
Group in the Spanish  Observatorio del Roque  de los Muchachos  of the
Instituto de Astrofisica de Canarias.}}

   \author{E. T. Harlaftis
          \inst{1}
          \and
          R. Baptista\inst{2}
	  \and
	  L. Morales-Rueda\inst{3}
	  \and
	  T. R. Marsh\inst{3}
	  \and
	  D. Steeghs\inst{4}
          }

   \offprints{E. T. Harlaftis}

   \institute{Institute of Space Applications and Remote Sensing,
National Observatory of Athens, P. O. Box 20048, Athens 118 10, Greece\\
              \email{ehh@space.noa.gr}
         \and
         Departamento de F\'{\i}sica, Universidade Federal de Santa 
Catarina, 
          Campus Trindade, 88040-900, Florian\'opolis, Brazil\\
             \email{bap@fsc.ufsc.br}
	 \and
	Department of Physics and Astronomy, Southampton University, 
	Southampton SO17 1BJ, UK \\
            \email{lmr@astro.soton.ac.uk, trm@astro.soton.ac.uk}
	\and
	 High Energy Astrophysics Division, Center for Astrophysics,
          MS-67, 60 Garden Street, Cambridge, MA 02138, USA\\
             \email{dsteeghs@head-cfa.cfa.harvard.edu}             
}

   \date{Received November 4, 2003; accepted December 27}

\abstract{We perform simulations in order to reveal  the effect of 
observational   and physical parameters   on  the reconstruction  of a
spiral structure  in an   accretion    disk, using eclipse     mapping
techniques.   We show that a  model spiral structure   is smeared to a
``butterfly''-shape   structure because of   the azimuthal   smoothing
effect of the technique.  We isolate  the effects of phase resolution,
signal-to-noise ratio  and accurate centering  of the  eclipse at zero
phase.  We further explore disk emissivity factors such as dilution of
the spiral  structure  by the  disk  light  and relative spiral   arm
difference.  We    conclude   that the    spiral  structure   can   be
satisfactorily recovered in  accretion  disk eclipse maps with   phase
resolution $|\Delta\phi|    \leq0.01$,   $S/N>25$ and    zero    phase
uncertainty $|\Delta\phi|  \leq0.005$,  assuming the two spiral arms have
similar brightness and  contribute $\geq 30$  \% to the total
disk light.  Under the light  of the performed simulations, we present
eclipse maps of  the IP Peg accretion  disk reconstructed from eclipse
light curves  of emission lines and  continuum  during the outburst of
August 1994, where spiral shocks were detected with the aid of Doppler
tomography (Morales-Rueda et al.  2000).  We discuss how the detection
of spirals shocks with eclipse mapping is improved with the use of
velocity-resolved eclipse light curves which do not include any 
contaminating low-velocity emission.
   }

   \maketitle
%

\section{Introduction : spiral shocks in cataclysmic variables}

Cataclysmic variables  (CVs) are binaries  where an accretion  disk is
formed around the white dwarf from  gas escaping a companion red dwarf
(Warner 1995).   Their  study is attractive   for understanding binary
formation and   evolution and the  physical  processes occuring in the
accretion  flow.  For example,   the semi-regular  radiation outbursts
from the accretion  disks around white dwarfs  in the  CV sub-class of
the dwarf   novae are thought  to  be driven  by a thermal instability
within the disk  (Hameury  2003 and references therein).   The optical
radiation pattern on the accretion disk is asymmetric and is dominated
either by the  ``bright spot'' caused by  the impact of the gas stream
on  the   accretion disk in quiescence   (Marsh  1989) or the ``spiral
structure''  raised  by tides  from the companion   star on  the outer
accretion disk during outburst (Steeghs 2001; Boffin 2001).

IP Pegasi is an eclipsing dwarf nova  and one of  the best studied CVs
thus providing a paradigm for  the accretion disk model. Spiral shocks
in its accretion disk have been found throughout the various stages of
an outburst (rise, maximum,  decline; Steeghs et al.  1997;  Harlaftis
et al.  1999; and Morales-Rueda et al.  2000, respectively), using the
Doppler   tomography  technique  (Marsh  \&  Horne   1988).  The  
phase-resolved spectra in Harlaftis et al.  (1999), which were used to
reconstruct the Doppler maps -  two-dimensional velocity maps - of the
spiral shocks, were also used to make eclipse  light curves in various
emission lines.  The light curves  (integrated intensity versus phase)
were then analysed using the eclipse mapping technique (Baptista 2001)
in  order  to reconstruct the  surface  brightness distribution of the
accretion disk and thus define  the spatial extent  of the spiral arms
(Baptista  et al.  2000).  Direct comparison of the Doppler (velocity)
and   eclipse (spatial) maps  showed that   the  outer disk  - largely
affected by the velocity field of the spiral shocks  - in IP Pegasi is
largely non-Keplerian and that the spiral  arms extend between 0.2-0.6
$R_{L_{1}}$, contributing between 16-30 \% of the total line flux.

Here, we address the  parameters  affecting the reconstruction of  the
spiral structure using the   eclipse mapping technique.  We  undertake
simulations in order to demonstrate the effect on the detectability of
spiral arms in situations with  an unfavorable combination of low S/N,
low time resolution, and the dilution of the shock  light by the disk.
Hence, we    define  the   observational  requirements    for   future
observations  of spiral shocks  using  the eclipse mapping  technique.
Finally, in the  light of the  simulations performed  we interpret the
eclipse maps   built from light curves of    IP Pegasi during outburst
(Morales-Rueda et al. 2000). This is  an epoch where the spiral shocks
have  been   excited  in the  accretion disk,    as shown   by Doppler
tomography.


\section{Simulations}

Doppler tomography  of IP  Peg has  clearly  revealed the disk  spiral
structure on a number of occassions and showed its strength in imaging
sub-structures within the accretion disk (Steeghs 2001).  However, the
disk spiral structure of IP Peg  is, apparently, not imaged as clearly
using the  eclipse  mapping technique (Baptista  et  al.  2000),  even
though the  eclipse  light curves are  from the  same outburst maximum
(November 1996).   There, the same  phase-resolved spectra show one of
the best examples of spiral shocks  in an accretion disk using Doppler
tomography as the reconstruction  technique (Harlaftis et  al.  1999).
As a  consequence, we have performed a  series of simulations in order
to better    understand  the structures reconstructed    with  eclipse
mapping.  The    reconstructions were obtained   using  the PRIDA code
(Baptista \& Steiner 1993), which computes an eclipse  map as an array
of 51x51  pixels, with a size of 2 $R_{L_{1}}$ and centred on the white dwarf.
Any additional uneclipsed component can also be modelled as a constant
flux. We  adopt the binary  geometry of Wood   \& Crawford (1986) with
($i, q$)=($81\degr$, 0.5), where $i$ is the inclination and $q$ is the
mass ratio (see also section 3.1).  For the reconstructions we adopted
the default  of limited azimuthal  smearing  (Rutten, van Paradijs  \&
Tinbergen 1992),  which  is better   suited for recovering  asymmetric
structures than the original default of full azimuthal smearing (e.g.,
Baptista 2001).    For  more details on  the   technique, see Baptista
(2001).    Using  a model   map  with a  symmetric  disk  and a spiral
structure, we test the limits of the technique in resolving the spiral
arms. The parameters  we  change are  either defined  by the observing
technique, geometrical constraints or physical processes :

\begin{itemize}
\item spiral structure orientation
\item phase resolution
\item signal-to-noise ratio
\item accurate zero phase of mid-eclipse
\item spiral structure dilution by the symmetric disk component
\item brightness difference between the two spiral arms
\end{itemize}

\begin{figure}
   \centering
   \includegraphics[angle=-90,width=8cm]{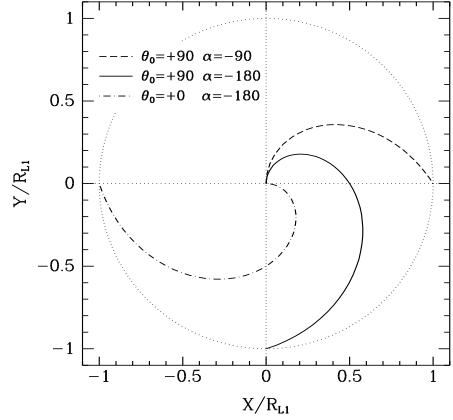}
\caption[]{The spiral arm definition using the parameters 
$\theta_{0}$, the  angle between  the  tangent  to the  spiral pattern
(extrapolated to the disk centre at $R_{\rm spiral}=0 ~R_{L_{1}}$) and the
line-of-centres, and $\alpha ~ (= -180
\degr)$, the opening  angle of the spiral, or the change in angle
($\theta   - \theta_{0}$)  it  takes   to  go from  $R_{\rm spiral}=0$  to
$R_{\rm spiral}= R_{L_{1}}$. The line-of-centres is the positive direction
of the x-axis and the angles are measured counter-clockwise.}
\label{spiral}
\end{figure}

\begin{figure*}
   \centering
   \includegraphics[angle=-90,width=\textwidth]{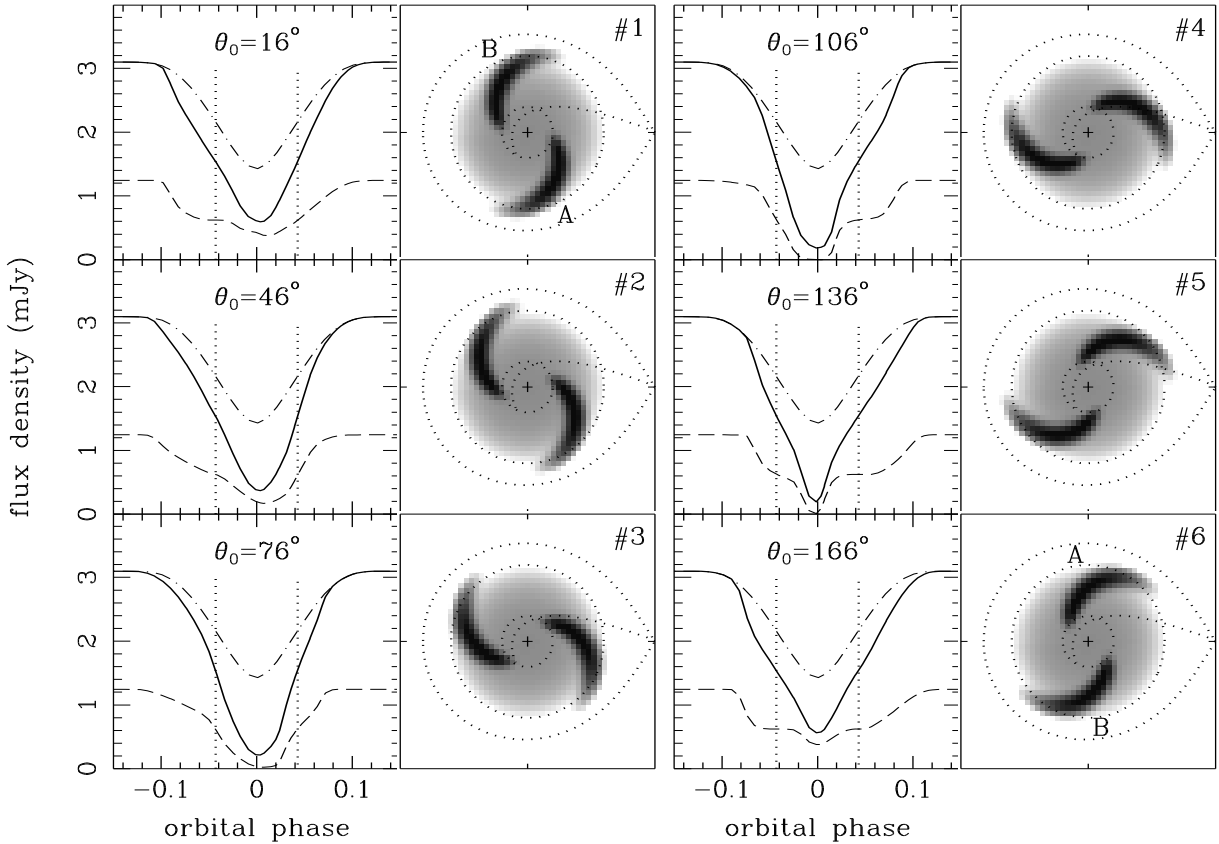}
\caption[]{A model accretion disk with a spiral structure at different
rotation angles  and the corresponding  total, disk, and  spiral light
curves.  The  model surface  brightness  distributions are shown in  a
logarithmic grayscale    (dark regions are   brighter).   Dotted lines
depict the primary Roche lobe, the gas  stream trajetory, and the disk
radius  at 0.2 and 0.6 $R_{L_{1}}$  (extent of observed spiral shocks;
Baptista et al.  2000).   A cross marks  the  disk centre.  The  stars
rotate counter-clockwise and  the  secondary is to  the  right of each
map.   In the   left-hand  panels, vertical  dotted   lines depict the
ingress/egress phases of the disk  centre (white dwarf).  The rotation
angle of the spiral   arm axis is  indicated  in each panel (see  also
text).    We adopt the model    with  an angle $\alpha=46\degr$  which
corresponds to the  one observed (Baptista  et al.  2000; Harlaftis et
al.  1999).  The model with  an angle $\theta_0=136\degr$ produces  an
eclipse shape which resembles that observed by Baptista et al. (2002).
The two spiral arms of equal brightness are labeled "A" and "B" in two
of the  panels.   The dashed lines in  the  left-hand panels  show the
contribution of the   two   spiral arms   to the light     curves; the
dot-dashed lines show the contribution of the symmetric disk component
to the  light curves; and the solid  lines show the total light curve.
For visualization purposes, the  curve of the symmetric disc component
is vertically shifted to match  the out-of-eclipse level of the  total
light curve.  Note  that the eclipse shape  is indeed affected  by the
rotation of the spiral structure. }
\label{simodel}
\end{figure*}

For the simulations, we  use as input a  model  map consisting of  two
spirals superimposed on  a symmetric disk. The  spirals have  a radial
width    of  0.08 $R_{L_{1}}$ and     a radial range  between 0.2--0.6
$R_{L_{1}}$, as found in Baptista  et al.  (2000).   The trace of  the
spiral is defined by,
\begin{center}
$R_{\rm spiral}(\theta) = ( \theta - \theta_{0} ) / \alpha$
\end{center}
where   $\theta_{0}$ is the angle   between the tangent  to the spiral
pattern  (extrapolated   to  the  disk  centre   at  $R_{\rm spiral}=0
~R_{L_{1}}$) and  the  line-of-centres and increases counter-clockwise
in steps  of $30\degr$,  using as a   reference the  observed rotation
angle  at  $46\degr$ (Baptista et al.    2000); $\theta$ is  the angle
between  the   tangent  to      the spiral pattern     at    $0<R_{\rm
spiral}<R_{L_{1}}$ and $\alpha ~ (= -180
\degr)$ is the opening  angle of the spiral, the change in angle
($\theta - \theta_{0}$) from $R_{\rm spiral}=0$ to $R_{\rm spiral}=0.6
~ R_{L_{1}}$.   Note that the  above  definition is for  linear spiral
arms  (see  Binney   \&   Tremaine   1988   for   logarithmic   spiral
arms). Fig. \ref{spiral} shows graphically  the spiral definition with
$\theta_{0}$  and $\alpha$ as  defined above.  Setting a small (large)
$\alpha$ results  in a tightly wound  (open) spiral arm. Our model map
in  Fig. \ref{simodel} has  an underlying symmetric disk containing 40
\% of the total  flux and a  flat brightness distribution which  drops
sharply at R=0.6  $R_{L_{1}}$  following  a $exp[-|(r-r_{0})/dr|^{3}]$
law.  The spirals  are also cut  off with the same radial  dependency,
both at the inner and outer radius.

\subsection{Spiral structure orientation ($\theta_{0}$)}

In Fig.  \ref{simodel},  we   present model  maps  of  spiral arms  at
different  rotation angles which we    use  as input images for    the
simulations,  so  that we compare the   predicted eclipse light curves
with the observed ones.  The dashed lines in the left hand panels show
the  contribution of  the two  spiral  arms  to the  light curves; the
dash-dotted  lines  show   the  contribution of   the   symmetric disk
component to  the light  curves;  and the solid  lines show  the total
light  curve.  The rotation angle  $\theta_{0}$ of the spiral arm main
axis   is  indicated in  each panel.    The   model surface brightness
distribution maps are  shown in a  logarithmic grayscale (dark regions
are brighter).  There, the dotted lines depict the primary Roche lobe,
the   gas stream trajectory,   and  the  disk  radii  at  0.2 and  0.6
$R_{L_{1}}$.  A  cross marks    the  disk centre.   The  stars  rotate
counter-clockwise  and the secondary is   to the right of each  panel.
The two spiral  arms, of equal brightness,  are labeled "A" and "B" in
two of the panels.  The change in the  shape of the eclipse is notable
as the  spiral arms rotate  from 16$\degr$  to 166$\degr$.   The wider
eclipse ingress  shifts  to a wider  egress  of the spiral  arms.  The
spiral  arm light curves  show best the change   in eclipse shape with
rotation angle.   Reconstructions  obtained from these   light  curves
confirm the  results of previous  simulations (Baptista et al.  2000),
namely, that the orientation and radial position of  the arms are well
recovered in  the eclipse maps.  The panels  present spiral models for
increasing  steps  of  $30\degr$ starting at   $16\degr$.  The eclipse
shape of the spiral model with rotation angle $46\degr$ resembles best
the  observed  eclipse  maps presented  here and   in Baptista et  al.
(2000). This   is clearer  in   the  next figure  which   presents the
simulation  results  with the  phase resolution  of the observed data.
The eclipse shape   of  the spiral model   with  $\theta_{0}=136\degr$
resembles the eclipse light curves of Baptista et al. (2002).

\begin{figure*}
   \centering
   \includegraphics[angle=-90,width=\textwidth]{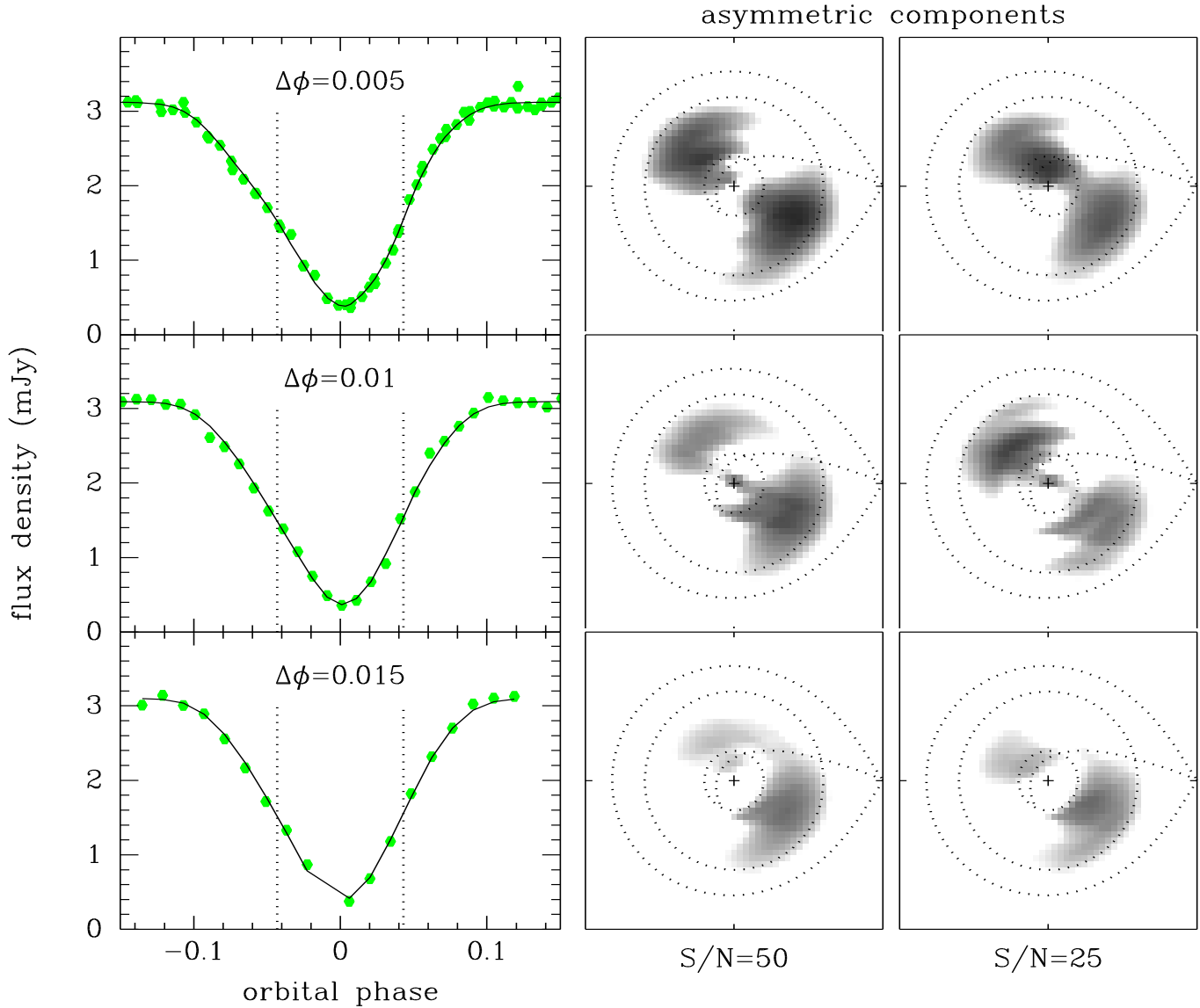}
\caption[]{The effect of the  phase resolution of  the  light curve on  the image
reconstructions.  The left-hand panels show data light-curves from the
second   panel of Fig.    \ref{simodel} for  three  phase resolutions:
$\Delta\phi$= 0.005,   0.01   and 0.015  cycles.    The  corresponding
asymmetric parts of the  eclipse maps for S/N=50  and 25 are presented
in logarithmic grayscale.}
\label{simphase} 
\end{figure*}

\subsection{Phase resolution ($\Delta\phi$) and S/N ratio}

Fig.  \ref{simphase} shows the effect of reducing the phase resolution
of the  light   curve on the    map reconstructions.  These  and   all
following simulations were  performed with  the adopted spiral   model
($\theta_{0}=46\degr$).  The left  hand panels show  data light curves
of S/N=50 for three phase  resolutions: $\Delta\phi$= 0.005, 0.010 and
0.015  cycles.  Gaussian  noise has  been added  in the light  curves.
Note that the  phase resolution of  $\Delta\phi = 0.010$ is similar to
that of the average  light curves extracted from  the blue spectra  of
the August 1994 outburst (see section  3).  The right-hand panels show
the asymmetric part  of   the resulting  eclipse maps  in  logarithmic
greyscale for S/N=50 and 25,  respectively.  The asymmetric  component
of each  eclipse map  is computed  as follows.  The  radial  intensity
profile is separated into 9 radial bins of  equal width.  We then sort
the intensities in each  bin, and fit a  smooth spline function to the
lower  quartile  of  the intensities  in each   bin.  The fit  is then
subtracted from each pixel  and the  resulting  map is written  as the
asymmetric   component.  We  practically  remove the   baseline of the
radial  profile, leaving  all  azimuthal structure  in  the asymmetric
map. We further set to zero all negative intensities in the asymmetric
map.

The quality of the reconstruction is sensitive to the phase resolution
of  the data.  The light  curve with the  highest phase resolution and
best S/N shows that the spiral structure is  considerably blurred to a
``butterfly''  pattern, as  a  consequence of  the  azimuthal smearing
effect of the    eclipse    mapping method. Poor      phase resolution
($\Delta\phi$=0.015) results in   spiral arms with  poorer definition,
since the  gradient change in the  eclipse light curve shape  -- which
testifies the presence of the spirals -- is poorly sampled.  The noise
masks  the ingress/egress of the spiral  arms (gradient changes in the
eclipse light curve  shape) and the eclipse  mapping method produces a
V-shaped fairly symmetric model light curve of constant ingress/egress
slope which will drive the solution to  a more symmetric configuration
by smearing the  spiral arms. Thus, the spirals  cannot be resolved on
the   map if the   light  curves have  low  S/N  (or  are affected  by
significant disk flickering).

\begin{figure}
   \centering
   \includegraphics[width=10cm]{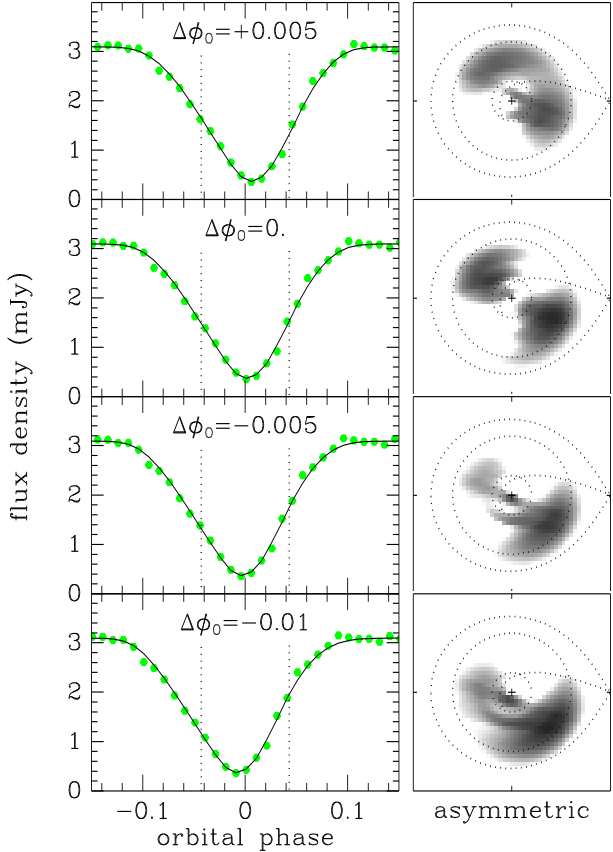}
\caption[]{The effects of accurate phasing of the mid-eclipse (phase offset 
$\Delta\phi_{0}$)   on    the   reconstructions.     The   value    of
$\Delta\phi_{0}$ is indicated  in each panel; the asymmetric component
of the corresponding eclipse maps are shown in the right panels in the
same logarithmic greyscale.   The notation is  similar to  that of the
other figures.  }
\label{simoffset} 
\end{figure}

\subsection{Accurate zero phase of mid-eclipse ($\Delta\phi_{0}$)}

In  Fig.   \ref{simoffset} we apply   additional phase  offsets to the
eclipse light curves in order to explore the effect on the spiral arms
of the uncertainty  on the exact zero phase  (0.006 cycles for IP Peg;
see section 3).  Small errors in the mid-eclipse, $\Delta\phi_{0}
\simeq 0.005$ cycles,  lead   to distortions  in the eclipse    map by
changing the azimuthal  position  and  extension of the  spiral  arms.
Positive (negative) phase offsets  displace both arms to  the trailing
(leading) side of the primary Roche lobe and also reduces the strength
of the  blue  (red)  arm,  thus  severely  distorting the  disk  light
distribution.  A  phase offset  of  -0.010   to the light   curves  is
sufficient to push  the light towards a  single  segment of  the outer
disk, in  fact  where  the  blue spiral arm   lies  at.  This exercise
emphasizes the importance of knowing the value of $\phi_{0}$ with good
accuracy.

\begin{figure*}
   \centering
   \includegraphics[width=\textwidth]{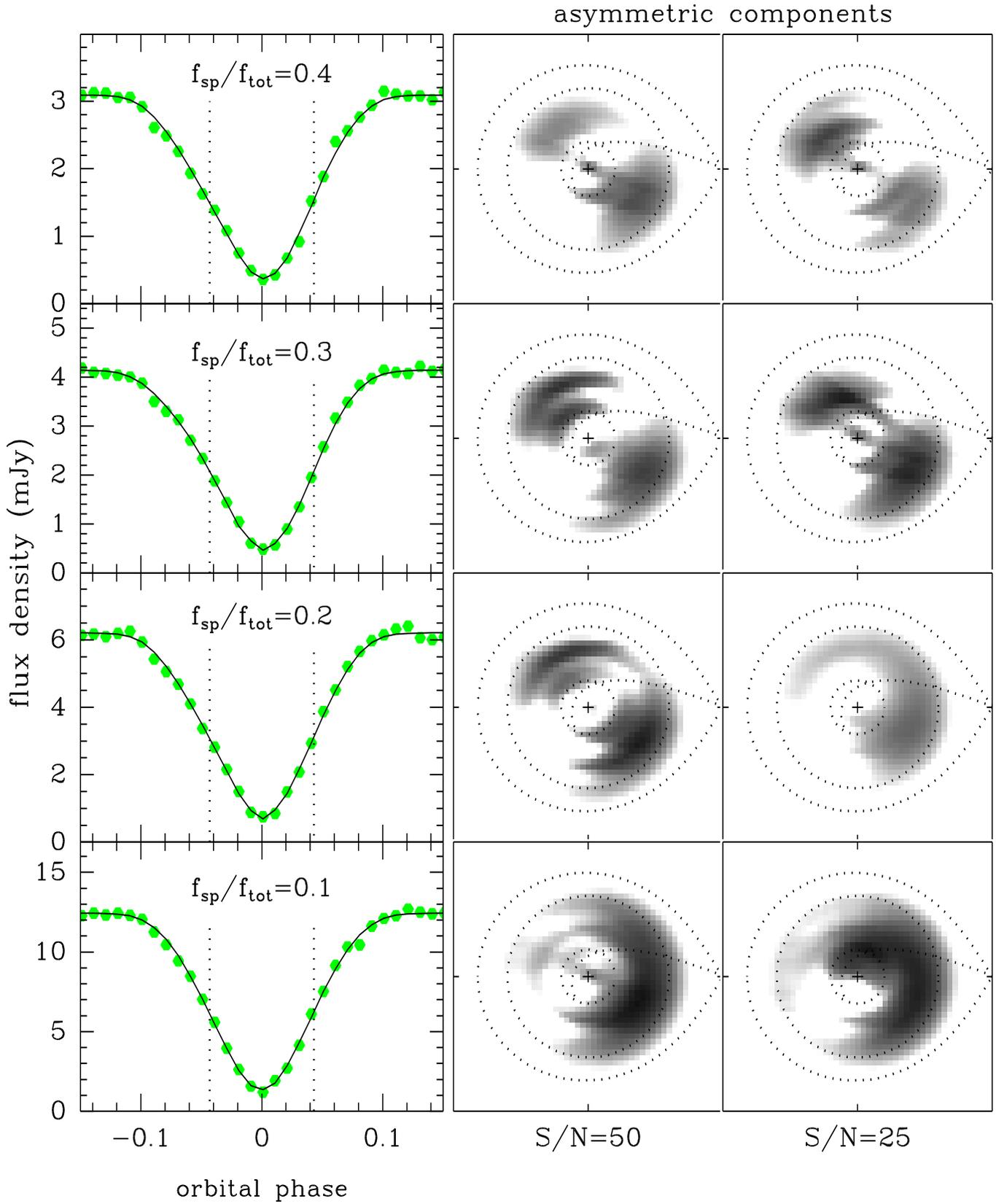}
\caption[]{The effects of dilution of the spiral structure light 
by a symmetric disk component on  the image reconstructions.  The left
hand panels show  model  light curves ($\Delta\phi=0.01$, S/N=50)  for
disk models ($\theta_{0}=46\degr$) in which the spirals contribute 40,
30, 20 and 10 per  cent of the total  flux.  The centre and right hand
panels show the asymmetric parts of the corresponding eclipse maps for
S/N=50  and 25, respectively.  When  the emission from the spiral arms
suffers large dilution (e.g.,  fractional contribution of 10 per cent)
the method  loses its ability to recover  the spiral arms  and returns
only an  asymmetric structure with a  ``crescent''  shape at the outer
disk and towards the $L_{1}$ point.}
\label{simdisk} 
\end{figure*}

\subsection{Spiral structure dilution by the symmetric disk component}

Fig.  \ref{simdisk} presents the test results  of the spiral structure
light  dilution caused by a  symmetric  disk component.  The left-hand
panels   show model light      curves  (with  phase   resolution    of
$\Delta\phi=0.01$ and S/N=50) for disk  models ($\theta = 46\degr$) in
which the spiral structure contributes  40, 30, 20  and 10 per cent of
the total flux.  The  right-hand panels show the asymmetric components
-  after subtracting the symmetric map  from the total intensity map -
of the corresponding eclipse maps for S/N=50 and 25, respectively.  At
small   fractional contributions  (10  \%), the   spiral structure  is
significantly diluted by the superimposed symmetric disk light and the
``butterfly''-like  pattern   (bottom    panels)   is smeared    to  a
``crescent''-like pattern at the outer disk  facing the $L_{1}$ point.
Provided that the spiral  structure contributes a significant fraction
to the total light ($f_{\rm spiral}/f_{\rm total} > 0.2$), the two-arm
structure is then reasonably recovered as a ``butterfly'' pattern.

Estimating the contribution of the spirals  from the ratio of the flux
in the asymmetric map component and that  of the total eclipse map may
lead to  systematically  underestimated fractional contribution.   The
reason fot  this is that  part of the  flux from  the spiral arms goes
into the symmetric component because of  the azimuthal smearing effect
of the eclipse mapping method.   The error increases for light  curves
of  lower phase resolution:  the fractional contribution of the spiral
arms is better recovered when the eclipse  light curve shape is better
sampled.  Using our  simulations,    we determine that the    inferred
fractional contribution of the spirals (40\%) may be underestimated by
a factor of 1.6-2.4 for light curves (S/N=25) with phase resolution of
$\Delta\phi = 0.010-0.015$ orbital cycles (as observed), respectively.
Light curves  with  S/N=50 and  $\Delta\phi =  0.005$ can recover  the
fractional contibution of the spiral structure  within 12\%.  At small
fractional contributions ($<$10\%), the spiral structure will start to
be  overestimated   (30-40 \%) by  the   technique.  The error  in the
fractional contribution of the spirals also depends  on the S/N of the
light  curve  but   is much smaller  than  that   caused  by low phase
resolution.  Thus, since we   estimated a fractional  contribution  of
0.15--0.3 in  Baptista et  al.   (2000), the  true contribution of the
spiral  arms in  those  eclipse maps may  be  in  the range  0.3--0.6.
Hence, the models with $f_{\rm spiral}/f_{\rm total}
\sim 0.3-0.4  $  should be a  good   approximation of the   real case.
In conclusion,  we can  only  place a lower  limit  to the  fractional
contribution of the spiral arms to the total light using eclipse maps.

\begin{figure*}
   \centering
   \includegraphics[angle=-90,width=\textwidth]{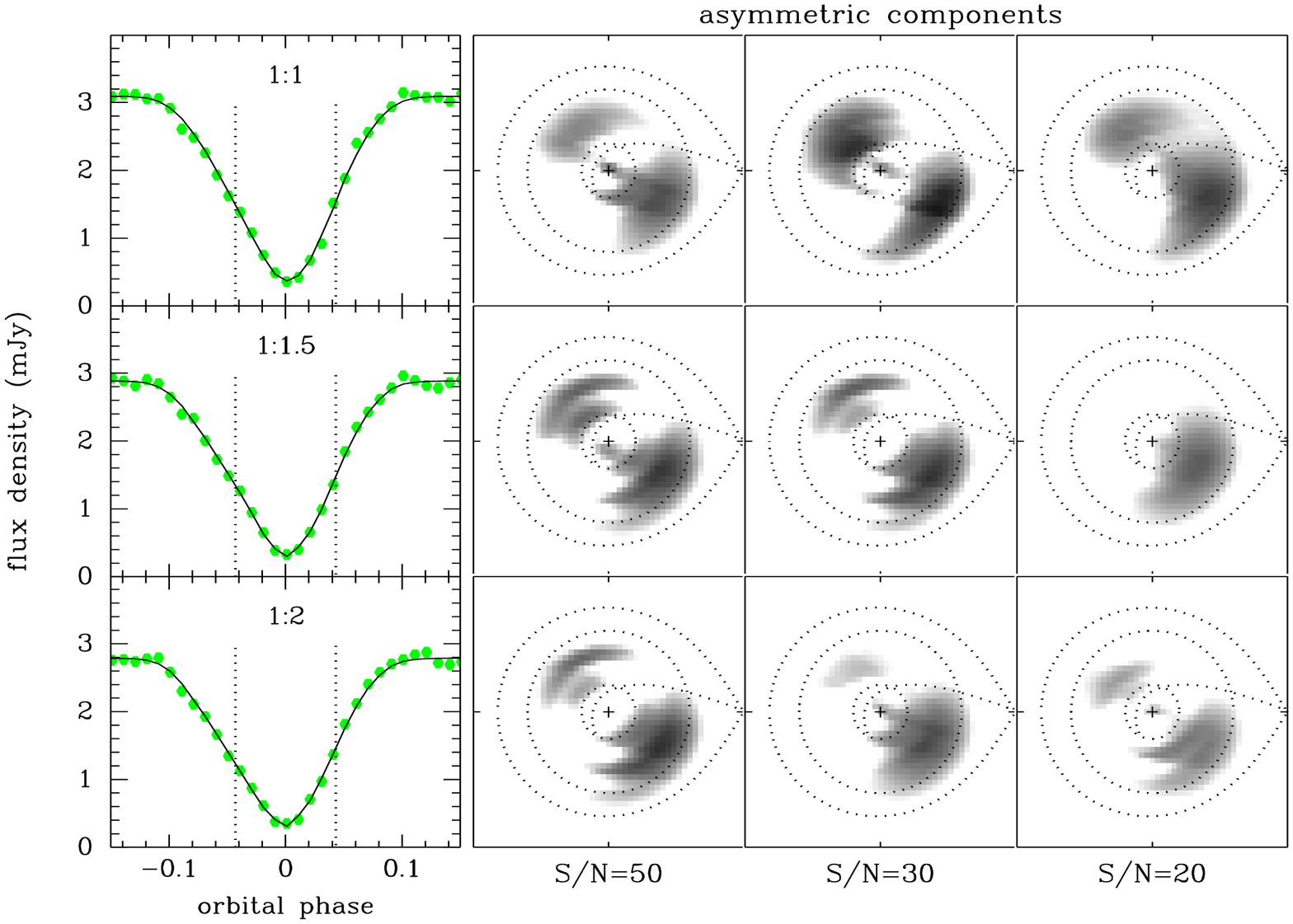}
\caption[]{The effect of brightness difference between the two spiral arms on the
image reconstructions. The left-hand panels show light curves (S/N=50)
for models in which the red and blue arms have equal brightness (top),
and models  in which the  red arm is fainter  than the blue arm in the
brightness  ratio  1:1.5 (middle)  and  1:2 (bottom).  The  right hand
panels show the corresponding eclipse  maps in log grayscale for three
different  S/N. Low  S/N data  (S/N$<$30) may lead   to a loss of  the
fainter spiral arm. Model arms of equal brightnesses are preserved
in the reconstruction map, regardless of the S/N.}
\label{simarmdiff} 
\end{figure*}

\subsection{Brightness difference between the two spiral arms}

Finally, we undertake a test with a spiral structure whose arms are of
unequal brightness.  This is driven by  the fact that observed Doppler
maps show an intensity difference between the  two  spiral
arms (Harlaftis et al.  1999; Steeghs 2001). Fig. \ref{simarmdiff}
presents the  effect of brightness  difference between  the two spiral
arms   on  the  map reconstructions.  We    build  model  light curves
($\Delta\phi=0.01$, S/N=50)  from  spiral  structure/disk  models,  as
described in Fig. \ref{simodel} in which the  red and blue arms have
equal brightness (1:1, top-left panel), or the red arm is fainter - as
observed - with  a   brightness  ratio  to  the  blue  arm   at  1:1.5
(middle-left panel) and finally at 1:2 (bottom-left panel).  The other
panels in   the figure show    the  corresponding  eclipse  maps    in
logarithmic grayscale for three different  S/N ratios (50, 30 and 20).
It is notable that both  arms  appear in the   map as a  ``butterfly''
pattern when  they have equal brightnesses  regardless of the S/N ratio.
However, the brightness difference between the  spiral arms at low S/N
ratio data ($<30$)  leads to loss of  the fainter spiral arm in the
image reconstruction.

\begin{table}
\begin{center}
\caption[]{Summary of Simulations}
	\label{}
\end{center}
	$$
\begin{array}{cccccc}
	\hline
        \noalign{\smallskip}
(f_{\rm spiral}/f_{\rm total})_{\rm model}  & (f_{\rm 
spiral}/f_{\rm total})_{\rm recon}& 
S/N 
& \Delta\phi &\\
\hline
	\noalign{\smallskip}
   0.40 &     0.17 &  25  &0.015 &{\rm Fig. 3}\\
   0.40 &     0.25 &  25  &0.010 & ``\\
   0.40 &     0.28 &  25  &0.005 & ``\\
   0.40 &     0.18 &  50  &0.015 & ``\\
   0.40 &     0.25 &  50  &0.010 & ``\\
   0.40 &     0.35 &  50  &0.005 & ``\\

   0.40 &     0.25 &  25  &0.010 &{\rm Fig. 5}\\
   0.30 &     0.32 &  25  &0.010 & ``\\
   0.20 &     0.13 &  25  &0.010 & ``\\
   0.10 &     0.14 &  25  &0.010 & ``\\
   0.40 &     0.25 &  50  &0.010 & ``\\
   0.30 &     0.27 &  50  &0.010 & ``\\
   0.20 &     0.19 &  50  &0.010 & ``\\
   0.10 &     0.13 &  50  &0.010 & ``\\
        \noalign{\smallskip}
        \hline
         \end{array}
     $$ 
\begin{list}{}{}
\item[] $(f_{\rm spiral}/f_{\rm total})_{\rm model}$: is the fractional 
contribution in the model eclipse map, 
\item[] $(f_{\rm spiral}/f_{\rm total})_{\rm recon}$: is the fraction 
measured by dividing the average flux in the asymmetric map by the average 
flux in the reconstructed eclipse map
\end{list}

\end{table}

\section{Spiral shock observations} 

\subsection{Eclipse maps of the August 1994 outburst}

Morales-Rueda  et al.    (2000)  present Doppler  tomography   showing
persistent  spiral shocks in the accretion  disk of  IP Peg during the
August  1994  outburst.  The  reader is  referred  to Morales-Rueda et
al.   (2000) for the  details of  the observations  and data reduction
procedure.  In  summary, 164 blue spectra  were obtained at a spectral
resolution of 100 km s$^{-1}$ at H$\beta$ and a time resolution of 220
seconds covering   the emission lines  H$\beta$,  H$\gamma$, He{\small
~II} 4686~\AA ~(4040--4983 ~\AA).  ~Here,  we extend their analysis in
order to include the eclipse maps for comparison  to the Doppler maps.
We combined the four eclipse light curves (E=  25158, 25159, 25164 and
25165) from  the  two consecutive  nights the  observations took place
(30/31 August 1994) to produce average eclipse light curves with phase
resolution of 0.010 cycles.   In particular, we extracted light curves
of the  blue continuum and of  the emission lines H$\gamma$, He{\small
~II}  4686~\AA, ~the Bowen blend  and the He{\small~I} 4472~\AA ~(Fig.
5 in Morales-Rueda et al. 2000) in order to  reconstruct the images of
the accretion disk.

We  processed the light   curves before applying  the  eclipse mapping
procedure so that the continuum level before  and after eclipse is the
same; otherwise, artifacts  are  raised.  These low-amplitude  orbital
variations were removed   from the light  curves  by fitting  a spline
function to the  phases outside eclipse, dividing  the light curve  by
the fitted spline, and  scaling the result  to the spline function  at
phase zero.   This procedure removes orbital  modulations, such as the
bright spot, with   only minor effects   on the  eclipse  shape itself
(e.g., Baptista   et al.  2000).   The other  factor  that affects the
eclipse mapping procedure (see simulation at Fig.  \ref{simoffset}) is
an accurate ephemeris for the eclipse minimum. Here, we use the linear
ephemeris by Wolf et  al.  (1993) after  correcting the phase zero for
mid-eclipse  rather than  the  white  dwarf  egress  (the white  dwarf
eclipse width is $\Delta\phi = 0.086$ cycles; Wood \& Crawford 1986)
\begin{center}
$T_{\rm mid}(HJD)= 2445615.4156 + 0.15820616 ~E$
\end{center}
where $T_{\rm mid}$ is the inferior conjunction of  the red dwarf.  IP
Peg shows cyclical orbital period changes of  amplitude 0.002 days (or
0.013 orbital cycles; Beekman  et al.  2002).  Furthemore, the scatter
of eclipse  timings for a given epoch  is also about 0.001 days (0.006
cycles).  Clearly, even  a sinusoidal ephemeris cannot predict eclipse
timings with precision  better than 0.01-0.02  cycles for IP Peg (Wolf
et al.   1993).   Therefore, the best  approach in   order to properly
phase the  light curves  is to center  the  white dwarf eclipse  using
contemporary quiescent eclipse timings.  The August 1994 data (eclipse
cycle $E
\simeq 25160$) is luckily bracketed by two HST sets of observations of
IP Peg in quiescence (Baptista  et al.  1994).  From these timings, we
infer that  the white dwarf mid-eclipse at  that  epoch occurred 0.008
orbital cycles before  the prediction of  the linear ephemeris of Wolf
et al  (1993).  Hence,  we added +0.008   cycles to the phases  of the
light   curves to make  phase  zero    coincident  with the   inferior
conjunction of the red dwarf.  The minimum of the eclipse in the lines is
displaced towards earlier phases  in comparison to the continuum light
curves.  Eclipse maps were then  made with three different geometries:
($i,   q$)=($81\degr$,   0.5)   (Wood   \&   Crawford    1986),   ($i,
q$)=($79.4\degr$, 0.58) (Marsh 1988)  and ($i, q$)=($86.5\degr$, 0.31)
(Beekman et  al.  2000), where $i$ is  the inclination and  $q$ is the
mass ratio.  These tests do not show any difference caused by a slight
change   in the  adopted orbital  parameters. Thus, we adopt in  the
following the geometry of Wood \& Crawford (1986).

\begin{figure*}
   \centering
   \includegraphics[width=\textwidth]{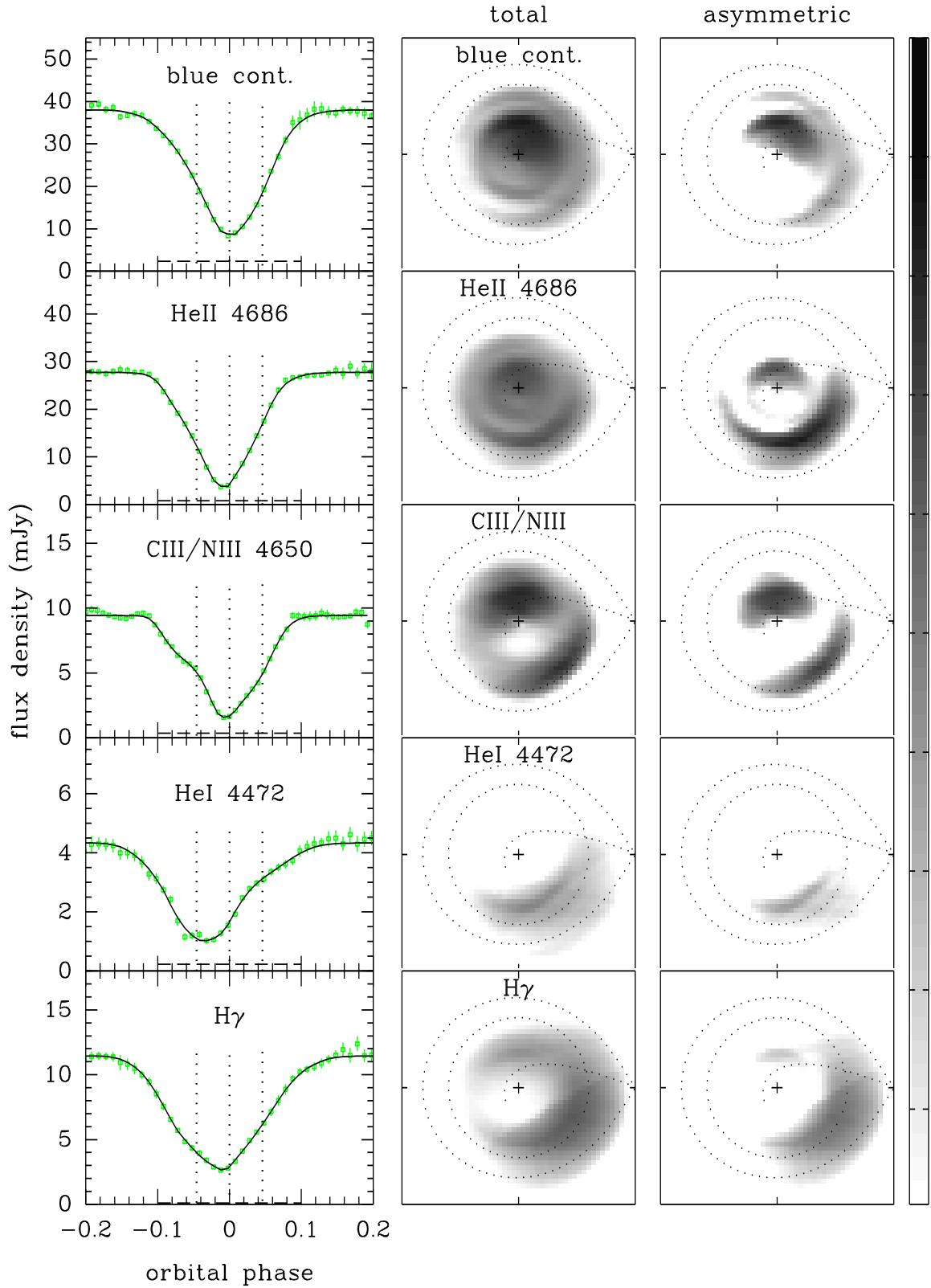}
\caption[]{The light curves, fits, eclipse maps and asymmetric  parts of the maps
of   the blue continuum  and  of  the  emission  lines He{\small  ~II}
4686~\AA, ~~Bowen blend, He{\small ~I} 4472~\AA and ~H$\gamma$ (August
1994 outburst; Morales-Rueda et al.  2000). The middle of ingress  and
egress of the disk centre are marked with vertical dotted lines in the
light curves and the fits to the data  are also plotted. Note that the
eclipse minimum varies in phase between the Balmer and helium emission
lines as well as with respect to the continuum minimum. The Roche lobe
of the white dwarf  is plotted together   with the disk radius at  0.6
R$_{L_{1}}$. The donor star lies to the right of the  disk map and the
ballistic trajectory of  the gas is  also marked.   The reconstruction
leads to asymmetric eclipse maps in which a brighter, outer asymmetric
arc is    seen together  with   a  fainter,   inner  arc  (logarithmic
grayscale).  The structure resembles that of a two-spiral arm only for
the Bowen and  He{\small  ~II}   maps.  Doppler  tomography   resolves
clearly    the  two   spiral  arms   using    the out-of-eclipse  data
(Morales-Rueda et al. 2000).}
\label{xfig01}
\end{figure*}

All maps  in Fig. \ref{xfig01} show an  asymmetric disk.  In  order to
resolve the asymmetric part of the disk we subtract the symmetric part
of the disk  and plot the residual on  the right panels of the figure.
The  continuum eclipse is  skewed towards the  egress and maps into an
asymmetric   disk  which  is brighter   towards   the  gas stream  and
$R_{L1}$. The Bowen  blend and  He{\small ~II} high-ionization   lines
show a relatively  narrow,  asymmetric eclipse  shape which maps  into
asymmetric brightness distributions  consistent with  spiral structure
and  similar to those in Baptista  et al.  (2000, see  their Fig.  2).
However,   the Balmer  and   He{\small   ~I} line  maps  are  markedly
different.  The strong asymmetric  Balmer emission is located close to
and even outside  the tidal radius,   with the maximum  emission at an
angle  of $-30\degr$ with respect  to the line-of-centres.  The centre
of the He{\small ~I} eclipse  occurs before that  of the Balmer lines,
which shifts  the asymmetric emission  towards a direction opposite to
the  gas stream (blue-shifted velocities  of the disk), and beyond the
tidal radius.  We apparently see only one of the arms in He{\small ~I}
and there is almost no evidence of the spirals  in the Balmer lines --
probably  because  their emission is   significantly  diluted by other
light  sources     in  these   lines  (see    following     section on
velocity-resolved light-curves).      The ``crescent''-shape   of  the
brightness   distribution of the  Balmer lines  is  reminiscent of the
distributions obtained  for simulations  where the spirals  contribute
only 10  per cent of  the light at  the  corresponding wavelength (see
Fig. \ref{simdisk}).  In comparison,  Morales-Rueda et al.  (2000) use
the out-of-eclipse spectra   from the same  orbital cycles  to  easily
resolve the spiral shocks using Doppler tomography.

\subsection{Eclipse maps from July 1998 outburst}

We  also     used   WHT  service    time    to obtain   phase-resolved
spectrophotometry with  the ISIS spectrograph  of an eclipse of IP Peg
during its July 1998 outburst decline, four days after the peak of the
outburst (18 July  1998).   The 23 blue  spectra (and  31 red spectra)
covered the range 4407.6--4847.7 \AA ~ (and 6310.7--6717.6 \AA) ~ at a
spectral resolution of 0.2 \AA~ per pixel (and 0.8 \AA, ~respectively)
and with an exposure of 120 seconds.   The spectra covered the eclipse
between 0.85 to 1.2 cycles at  a phase resolution  of 0.011 cycles for
the   red   spectra and 0.014  cycles   for  the   blue  spectra.  The
reconstructed eclipse maps   show similar distributions to the  August
1994 maps, but with inferior quality  which is most  likely due to the
poorer  phase resolution and   the later  outburst  stage   (closer to
quiescence which should   result in a  fainter  spiral  structure). In
summary, the eclipse maps are consistent with the reconstructions from
the   August 1994 data  (previous  section) and  the  August 1996 data
(Baptista et al.  2000).

\subsection{Velocity-resolved eclipse maps : Revealing the spiral shocks}

\begin{figure*}
   \centering
   \includegraphics[width=14cm]{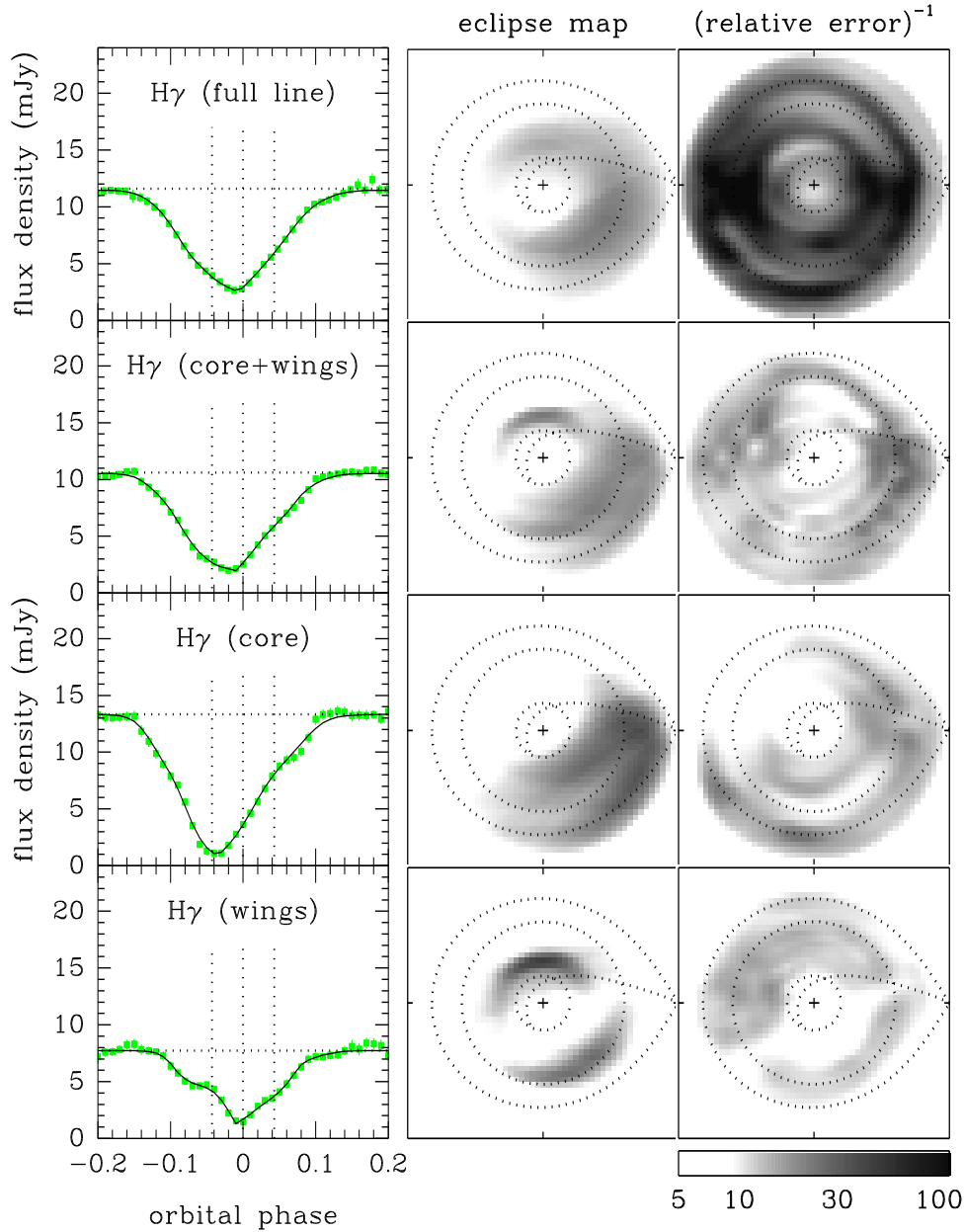}
\caption[]{Velocity-resolved (core/wings) eclipse   light curves can reveal with
clarity the spiral structure.  The core of the H$\gamma$ line profiles
($\pm250$  km  s$^{-1}$) during  eclipse    result in a  reconstructed
emission distribution towards the $L_{1}$.  The wings of the H$\gamma$
line  profiles ($\pm700-1300$ km  s$^{-1}$) during eclipse result in a
reconstructed  emission   distribution  consistent  with  a spiral-arm
structure. The earlier mid-eclipse caused by the low-velocity emission
component  results  in the  distorted, ``crescent''-like  eclipse map.
Lines, lobes and circles as in Fig. \ref{xfig01}. The right-hand panels
present S/N ratio maps, made using Monte-Carlo simulations (pixels with
S/N$\leq~5$ are painted white}.
\label{xfig05} 
\end{figure*}

We investigate here velocity-resolved eclipse maps since  it is a well
established  fact that there are  non-disk  emission components to the
total  line fluxes, for  example from  the  donor star during outburst
(Harlaftis 1999).   During outburst, another contaminating line source
to  the disk emission is  a low-velocity emission component (Harlaftis
et  al.  1999; Steeghs et  al.  1996). Since the  spiral  arms and the
secondary  star contribute to  the  line profile at  different Doppler
velocities,  it is  possible  to separate   the contribution of  these
distinct light  sources by   suitable extraction of  velocity-resolved
eclipse    light  curves.    Baptista    et   al.   (2002)   performed
velocity-resolved  eclipse  mapping    to separate    the He{\small~I}
emission from each of the spiral arms.  Here, we use this technique as
a  useful   tool   in order to     separate the  contribution   of the
low-emission components  (secondary star,   centre-of-binary emission;
Harlaftis et  al.  1999) from  the disk  emission and,  thus, make the
spirals more easily detectable in the  line eclipse maps. For example,
we can  extract eclipse light  curves  from the blue (red)  velocities
alone. The application of the eclipse mapping technique results in the
blue  (red) part of the  accretion disk (iso-velocity  surfaces on the
disk due to Keplerian rotation).  The sum of the blue and red parts is
indeed, as expected, consistent  with the total  image as built by the
velocity-integrated light curve.  After this test, we proceed with the
aim to isolate the inner/outer disk by extracting eclipse light curves
from different velocity ranges.  For this purpose, we extracted Balmer
and He{\small~I}  light  curves for the   line core (-250   to +250 km
s$^{-1}$) and for the  line wings (-1300 to  -700 and +700 to +1300 km
s$^{-1}$).  The eclipse light curves from the core of the Balmer lines
show a  broad   eclipse $\Delta\phi>0.2$ whose  minimum   is displaced
towards earlier phases with respect to the continuum mid-eclipse.  The
eclipse light curves from the wings of the Balmer lines are centred at
phase zero and show the same asymmetric eclipse shape to the He{\small
~II} and Bowen blend light curves with a clear step during ingress.

Fig. \ref{xfig05}  illustrates  the effects  of  low-velocity emission
components in diluting the spiral  structure embedded in the accretion
disk.  The figure   presents the results   of  eclipse mapping of  the
H$\gamma$ line, from top to  bottom, for the velocity-integrated total
profiles (full line), the sum of the core/wings velocity regions,  the core of the
profiles, and the wings of the  profiles.  The reconstruction from the
``core'' light curve has a broad asymmetry, towards the $L_{1}$ point,
in the outer regions  of the eclipse  map.   The eclipse map from  the
``wings'' light curve clearly shows  the two arc-shaped spiral arms at
the same orientation and azimuthal extent as are the ones in the Bowen
blend and He{\small   ~II}  eclipse maps   (Fig.   \ref{xfig01}).  The
``core/wings'' light curve (and  corresponding eclipse map) is similar
to  the  velocity-integrated light curve   (and eclipse map).  Similar
results are found for the H$\beta$ and He{\small~I} lines.  This phase
offset from mid-eclipse and the  resulting asymmetries in the  eclipse
maps are seen in the Balmer and He{\small~I}  lines (which also have a
substantial contribution from the secondary star at outburst), but are
not present in the He{\small~  II} and Bowen   blend light curves  and
eclipse maps (the contribution of the secondary star to these lines is
weak   or negligible).  The   eclipse mapping  technique,  in order to
account for the wide  and offset eclipse, tries  to build a brightness
distribution  in the disc regions  being ``eclipsed'', irrespective if
the physical origin lies on the secondary star or a disk wind, thereby
producing a fake, broad asymmetry in the outer  regions of the eclipse
map displaced towards the $L_{1}$ point.

On the right-hand panels, we present the  S/N ratio of the maps built.
For each   eclipse light  curve, a   set of 15  artificial curves  was
generated, in  which the data  points  were independently and randomly
varied according to a Gaussian   distribution with standard  deviation
equal to the uncertainty at  that point.   The artificial curves  were
fitted using the eclipse mapping technique to produce a set of eclipse
maps, based on random  data.  These, then,  were combined to produce a
``residuals'' map by taking the pixel-to-pixel standard deviation with
respect  to  the  mean    intensity.   This  yields   the  statistical
uncertainty at each pixel.  The uncertainty  of the standard deviation
for N=15 is $<$  20\% which is  sufficient for an illustration of  the
confidence limits of the eclipse map (more than  an order of magnitude
of Monte    Carlo  simulations are needed  in    order  to  reduce the
uncertainty of  the standard deviation down  to 7  \%).  The result of
dividing the eclipse maps of the observed data (or the mean map of the
Monte Carlo   simulations) by the ``residuals''  map  is shown  in the
right-hand panels of Fig. \ref{xfig05}, as maps of  the inverse of the
relative error  or   S/N  ratio  maps.   Darker regions   have smaller
uncertainties (greyscale is  such that  pixels  with S/N $\leq 5$  are
shown in  white color, while all   pixels with S/N$>$100 are  shown in
black color). The spiral   arms  in the  H$\gamma$ ``wing''   map have
typical S/N = 12--20.  The H$\gamma$ map (full  line) has the best S/N
because it was integrated over the whole line profile.

There may be various sources of low-velocity emission in the system to
produce the above  effects on the  eclipse  shape.  One such  emission
component of  unknown  origin  but close  to  the  binary centre   and
slightly towards the  gas stream has been seen  at outburst maximum in
IP Peg (Harlaftis  et al.  1999; Steeghs et   al. 1996).  Wind-related
emission  has also been suggested by  Baptista et al.  (1995) in order
to accommodate  excess light that   was not eclipsed  by the companion
star.  The   velocity of  the   red star, as  measured  from different
investigators, is also low either  at 280, or  300 or 330 km  s$^{-1}$
(Wolf  et al.  1998;   Marsh \& Horne  1990;   Beekman  et al.   2000,
respectively).  Irradiation of  the hemisphere of  the star facing the
white dwarf during outburst  causes a significant contribution to  the
Balmer and He{\small ~I} total  emission (Morales-Rueda et al.  2000),
and even  in  He{\small ~II}  at  outburst  maximum (Harlaftis  et al.
1999).  The implication is that if the emission pattern from the Roche
lobe is not  symmetric (possibly due to  shielding by the  bright spot
and gas stream), the centre of the  self-eclipse of the secondary star
line  emission happens earlier  than  the disc eclipse  at phase zero.
However, the  data from Morales-Rueda  et al.  (2000)  do not show any
solid  evidence for an  asymmetric emission distribution  on the Roche
lobe of the companion star.

\section{Discussion}

We have performed a thorough investigation to determine the conditions
under which   it is possible to detect   spiral structure in accretion
disks  of cataclysmic  variables, and  then  compared the results with
observed data.  The simulations (see Table  1 for a summary) show that
the phase resolution   combined with  the  S/N ratio  are  critical in
resolving  the features caused by the   spiral shocks in the eclipse
light curve  shape.  Indeed, poor phase resolution ($\Delta\phi>0.01$)
and low S/N ratio   ($<25$) can distort  the spiral   structure beyond
visible recognition   (bottom  panels   in Fig.   \ref{simphase}).  In
addition, the eclipse    mapping  technique assumes a uniform,    flat
accretion disk and this smears the spiral pattern into a ``butterfly''
pattern (top panels in Fig.  \ref{simphase}).

Aside for the above observational parameters (time resolution and S/N)
and technical effects (maximum entropy  method and model assumptions),
the   emissivity  parameters  of  the spiral     structure are equally
important     in order  to  reveal  it    using   the eclipse  mapping
technique. Dilution by the accretion disk  light makes it difficult to
resolve the spiral  structure.  The dilution  of light ($>$ 70\%)  can
distort, through  the  eclipse mapping technique,  the  smeared spiral
structure (``butterfly''  pattern) to  a  ``crescent''-like  structure
(bottom panels in Fig. \ref{simdisk}).  In addition to the above, the
comparative   emissivity between the two   spiral arms may  not be the
same, as indeed  observations indicate (``red''-arm weaker;  Harlaftis
et  al.  1999, Steeghs   2001).    This also  contributes to   further
attenuation  of  the  reconstructed  ``red'' arm   (bottom  panels  in
Fig. \ref{simarmdiff}). Poor phase resolution  itself may  result in
loss of the red spiral arm even if its  brightness is equal to that of
the  other arm.  Having good  phase resolution  is even more important
than having high  S/N light curves.   The observed maps presented here
from the early  decline of the August 1994  outburst (but also similar
results from spectra at late decline of the July 1998 outburst) can be
more  easily  read given the  simulations.   They are  consistent with
reconstructing the spiral structure as a  smeared and distorted spiral
structure, and   frequently one-arm pattern   in the  accretion  disk.
Similarly,  the maps presented by  Baptista et al.  (2001) at outburst
maximum and by Baptista, Haswell \&  Thomas (2002) at outburst decline
can be interpreted  in the same  fashion.  The WHT observations  (July
1998 outburst) show an  application for the case  where the data  does
not match the requirements put forward by the simulations, and the INT
observations (August 1994 outburst)  is an example of  a data set that
meets the observational requirements.

The presence  of spiral arms are inferred by small asymmetries in the
shape of the  eclipse light curve. When the  noise in  the light curve
becomes comparable to   the depth  of  these  asymmetries, the  method
looses the     ability to distinguish   (or  disentangle)   the spiral
structure from the  noise.  Also, the  characterization of the  spiral
structure   in  the   eclipse maps   demands  that  the  corresponding
asymmetries in eclipse shape are properly sampled in phase.  For light
curves of  poor   phase resolution  (or, more specifically,   when the
number of data points sampling the asymmetry  caused by the spiral arm
is small) it  is  hard to distinguish  the true  asymmetry from random
deviations in the  eclipse  profile caused  by noise.   These  effects
become more important if the contribution of the spirals is diluted by
other sources such as a symmetric disc component, because the relative
depth  (and significance) of   the asymmetries decrease for increasing
brightness of the additional source.

In the absence of dilution by  other light sources,  as with the Bowen
blend  light curve (Fig.   \ref{xfig01}), the spiral structures may be
well recovered even   for  light  curves  of relatively  poor    phase
resolution and incomplete phase coverage  such as those of Baptista et
al.  (2000;   see their  Fig.2).  In   the  presence of  light sources
diluting the spiral  structure,  optimal extraction of   eclipse light
curves in narrow passbands  matching the velocity  range of the spiral
arms  is an  interesting way to  optimize  the ability  of the eclipse
mapping method  in recovering  these  asymmetric  structures.  Indeed,
this   is the case  with  the eclipse light curves  of  the Balmer and
helium lines (Fig. \ref{xfig05}) where an unknown source of asymmetric
light  shifts  the mid-eclipse   of the  emission  lines resulting  in
distorted eclipse maps (``crescent''-shape or appearance of a ``single
spiral-arm'')  with the  surface  brightness  distribution towards the
companion  star. Suitable   velocity-range  selection from  the   line
profiles can avoid any contamination to  the spiral structure and thus
resolve  clearly the spiral structure  in the accretion disk.  Such is
the case, for example, with the eclipse light curve extracted from the
``wings''  of  the   emission-line    H$\gamma$   profiles and     the
corresponding eclipse map in Fig.
\ref{xfig05}. Given  the many  parameters  involved,  it is  not  a
surprise that  the spiral structure  was first discovered with the use
of Doppler tomography.  Doppler  tomography  defines very clearly  the
spirals  since   it provides   more   constraints from  phase-resolved
spectroscopy  (velocity, intensity, phase).

We  conclude that  the  eclipse  mapping method   is able  to properly
reproduce   the spiral structure provided   that  the observations are
tuned  (S/N$>$25, and $\Delta\phi$=0.01   or a time  resolution  of at
least 2 minutes for IP Pegasi), the data analysis is careful (avoiding
any contaminating   low-velocity emission   by selection   of suitable
velocity-resolved  light curves), and  the physical characteristics of
the spiral arms are  sufficient for  detection (contribution of  $\geq
30$ \% of the total disk light and minor brightness difference between
the two spiral arms).

The analysis we  have  undertaken using  eclipse  maps from   real and
synthetic data demonstrate how future observation runs can be tailored
in  order to  deduce in  full detail  the significance  of  the spiral
structure in  fast-rotating cataclysmic variable disks.   The presence
of  spiral structure suggests that  the application of eclipse mapping
techniques could  benefit from the use of  a default  model disk which
smears structures along spiral patterns instead of the usual azimuthal
direction (Harlaftis et al.,2004, in preparation).

\begin{acknowledgements}
The INT and WHT telescopes are  operated on the island of La
Palma by the Isaac Newton Group in  the Spanish Observatorio del Roque
de los Muchachos of the Instituto de  Astrofisica de Canarias. The WHT
spectra of IP  Pegasi were obtained  through the SERVICE  programme of
the Isaac Newton  Group. We thank an anonymous referee for his comments.
\end{acknowledgements}

\label{lastpage}

\end{document}